# Critical paths in a metapopulation model of H1N1: Efficiently delaying influenza spreading through flight cancellation

**Disease spreading through human travel networks has been a topic of great interest in recent years, as witnessed during outbreaks of influenza A (H1N1) or SARS pandemics. One way to stop spreading over the airline network are travel restrictions for major airports or network hubs based on the total number of passengers of an airport. Here, we test alternative strategies using edge removal, cancelling targeted flight connections rather than restricting traffic for network hubs, for controlling spreading over the airline network. We employ a SEIR metapopulation model that takes into account the population of cities, simulates infection within cities and across the network of the top 500 airports, and tests different flight cancellation methods for limiting the course of infection. The time required to spread an infection globally, as simulated by a stochastic global spreading model was used to rank the candidate control strategies. The model includes both local spreading dynamics at the level of populations and long-range connectivity obtained from real global airline travel data. Simulated spreading in this network showed that spreading infected 37% less individuals after cancelling a quarter of flight connections between cities, as selected by betweenness centrality. The alternative strategy of closing down whole airports causing the same number of cancelled connections only reduced infections by 18%. In conclusion, selecting highly ranked single connections between cities for cancellation was more effective, resulting in fewer individuals infected with influenza, compared to shutting down whole airports. It is also a more efficient strategy, affecting fewer passengers while producing the same reduction in infections.**

Contents

more

# Introduction

Complex networks are pervasive and underlie almost all aspects of life. They appear at different scales and paradigms, from metabolic networks, the structural correlates of brain function, the threads of our social fabric and to the larger scale making cultures and businesses come together through global travel and communication [1][2][3][4][5][6].

Recently, these systems have been modelled and studied using network science tools giving us new insight in fields such as sociology, epidemics, systems biology and neuroscience. Typically components such as persons, cities, proteins or brain regions are represented as nodes and connections between components as edges [6][7].

Many of these networks can be categorised by their common properties. Two properties relevant to spreading phenomena are the modular and scale-free organization of real-world networks. Modular network consist of several modules with relatively many connections within modules but few connections between modules. Scale-free networks with highly connected nodes (hubs) where the probability of a node having k edges follows a power law $k^{-\gamma}$ [8][9]. It is possible for a network to show both scale-free and modular properties, however the two features may also appear independently. The worldwide airline network observed in this study was found to be both scale-free and modular [10].

Spreading in networks is a general topic ranging from communication over the Internet [11][12], phenomena in biological networks [13], or the spreading of diseases within populations [14]. Scale-free properties of airline networks are of interest in relation to the error and attack tolerance of these networks [5][15]. For scale-free networks, the selective removal of hubs produced a much greater impact on structural network integrity, as measured through increases in shortest-path lengths, than simply removing randomly selected nodes [15]. Structural network integrity can also be influenced by partially inactivating specific connections (edges) between nodes [16][17][18]. Dynamical processes such as disease spreading over heterogeneous networks was also shown to be impeded by targeting the hubs [19][20], with similar findings for highest traffic airports in the case of SARS epidemic spreading [5].

In contrast to predictions of scale-free models, recent studies of the airline network [21] demonstrated that the structural cohesiveness of the airline network did not arise from the high degree nodes, but it was in fact due to the particular community structure which meant some of the lesser connected airports had a more central role (indicated by an higher betweenness centrality, the ratio of all-pairs shortest paths crossing each node). Here we expand on this finding further by considering a range of centrality measures for individual connections between cities, show that their targeted removal can improve on existing control strategies [5] for controlling influenza spreading and finally discuss the effect of the community structure on this control.

To demonstrate the impact on influenza spreading caused by topological changes to the airline network, we run simulations using a stochastic metapopulation model of influenza [22][23] where the worldwide network of commercial flights is used as the path for infected individuals traveling between cities (see Fig. 1A with Mexico City as starting node of an outbreak). For this, we observe individuals within cities that contain one of the 500 most frequently used airports worldwide (based on annual total passenger number). Individuals within the model can be susceptible (S), infected (I), or removed (R). The number of infected individuals depends on the population of each city and the volume of traffic over airline connections between cities. Note that the time course of disease spreading will also be influence by seasonality [23]; however, only spreading in one season was tested here.

# Figure 1

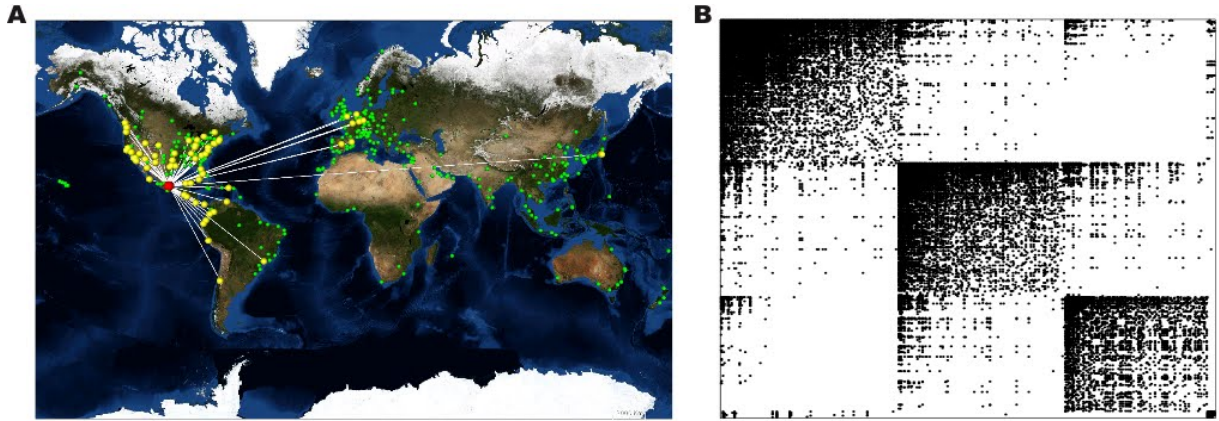

(A) Spreading over the airline network with Mexico City as starting node (red). Nodes in yellow are directly connected whereas nodes in green are airports not directly linked to the starting point. (B) Connectivity of the airline network showing four clusters. A dot in the adjacency matrix indicates the presence of a connection between two cities.

The simulated epidemic starts in 1 July 2007 from a single city, Mexico City in our case, and its evolution over the following year is recorded. We then consider the number of days necessary for the epidemic to reach its peak as well as the maximum number of infected individuals (Fig. 2A). This procedure is then repeated following the removal of a percentage of connections ranked as by a range of distinct measures such as edge betweenness centrality, Jaccard coefficient or difference and product of node degrees. Finally we also test the effect of shutting down the most highly connected airports (hubs) up to the same level of cancelled connections.

Comparing single edge removal strategies against the previously proposed shutdown of whole nodes (airports) we find that removing selected edges has a greater impact on the spreading of influenza with a significantly smaller loss of connectivity between cities. For the global airline network only a smaller set of flights routes between cities would need to be stopped instead of cancelling all the flights from a set of airports to get the same reduction in spreading.

In addition as demonstrated in [21] for structural cohesiveness and in [24] regarding dynamical epidemic spreading, it is the community structure and not the degree distribution that plays a critical role in facilitating spreading. Our method of slowing down spreading by removing critical connections is efficient as it targets links between such communities.

Concerning the computational complexity, whereas some strategies are computationally costly for large or rapidly evolving networks, several edge removal strategies are as fast as hub removal while still offering much better spreading control.

Note that whereas we observed similar strategies in an earlier study [25], the current work includes the following changes: First, simulations run at the level of individuals rather than simulating whether the disease has reached ('infected') airports. Second, the spreading between cities, over the airline network, now depends on the number of seats in airline connections between cities. This gives a much more realistic estimate of the actual spreading pattern as not only the existence of a flight connection but the specific number of passengers that flow over that link is taken into account. Third, the previous study used an SI model that is suitable for early stages of epidemic spreading. However, in this study we use an SIR model that allows us to observe the time course of influenza spreading up to one year after the initial outbreak.

# Results

For the network used in the study, the top 500 cities worldwide with the highest traffic airports became the nodes and an edge connects two of such nodes if there is at least one scheduled passenger flight between them. Edges are then weighted by the daily average passenger capacity for that route. Spreading in this network can then show how a disease outbreak, e.g. H1N1 or SARS influenza, can spread around the world [5][23].

As in previous studies [5][26], we have used a similar methodology [22] where one city is the starting point for the epidemic and air travel between such cities offers the only transmission path for an infectious disease to spread between them. Due to the relevance of the recent H1N1 (Influenza A) epidemic we have used Mexico city to be the epidemic starting point of our simulations.

Spreading simulations starting in Mexico City with 100 exposed individuals were summarised by $N_{infectious}$ the greatest number of infected individuals that were infectious at any time during the epidemic. Spreading control strategies were evaluated by removing up to 25% of the flight routes and measuring the resulting decrease in $N_{infectious}$ (see Fig. 2A and Methods). Measures based on edge betweenness and Jaccard coefficient were the two best predictors of critical edges (Fig. 1A). Among the top intercontinental connections identified by betweenness centrality are flights from Sao Paulo (Brazil) to Beijing (China), Sapporo (Japan) to New York (USA) and Montevideo (Uruguay) to Paris (France). After removing a quarter of all edges, both strategies showed a decrease in infected population of 37% for edge betweenness centrality and 23% for the Jaccard coefficient, compared to only 18% for the hub removal strategy.

# Figure 2

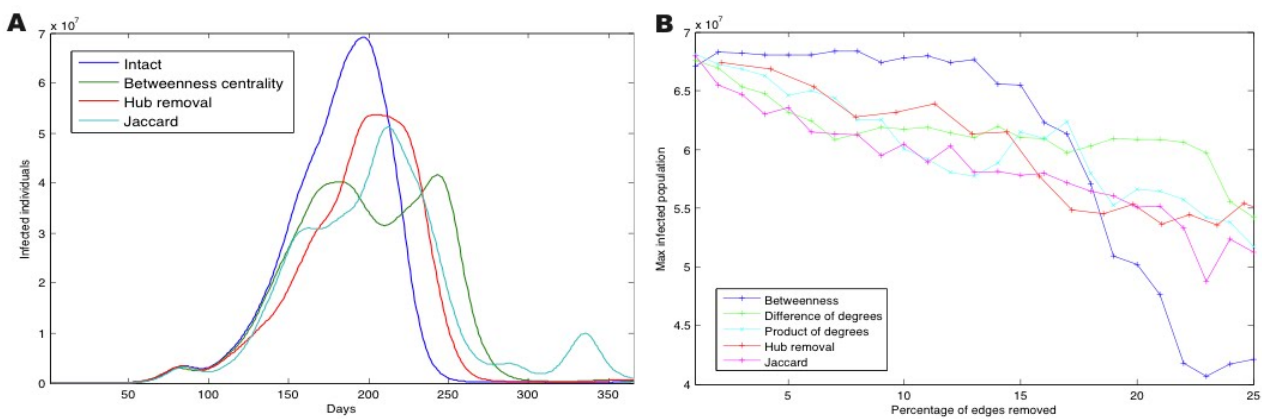

(A) Influenza spreading for Mexico City as starting node, measured by the number of infected individuals over time on the intact network (blue) and after removing 25% of edges by hub removal (red) or edge betweenness (green). (B) Maximum infected population following sequential edge elimination by betweenness centrality, Jaccard coefficient, difference and product of degrees and hub removal (see Methods).

Whereas in [23] a control strategy based on travel restrictions found that travel would need to be cut by 95% to significantly reduce the number of infected population, we observed that by removing connections ranked by edge betweenness this reduction to appeared after 18% of flight routes were cancelled (see Fig. 2B).

To understand the underlying mechanism of these results we produced two rewired versions of the original network: one version preserved the degree distribution alone while another preserved both the latter and also the original community structure.

Applying the same spreading simulations on these rewired versions of the network showed that only on networks that preserved the original's community structure did we observe a significant reduction in infections when removing edges (see Fig. 3) connecting nodes ranked by Jaccard coefficient. For the 25% restriction level considered, betweenness centrality was the best measure even when no communities were present, offering a 41% reduction in infected cases in both types of network.

# Figure 3

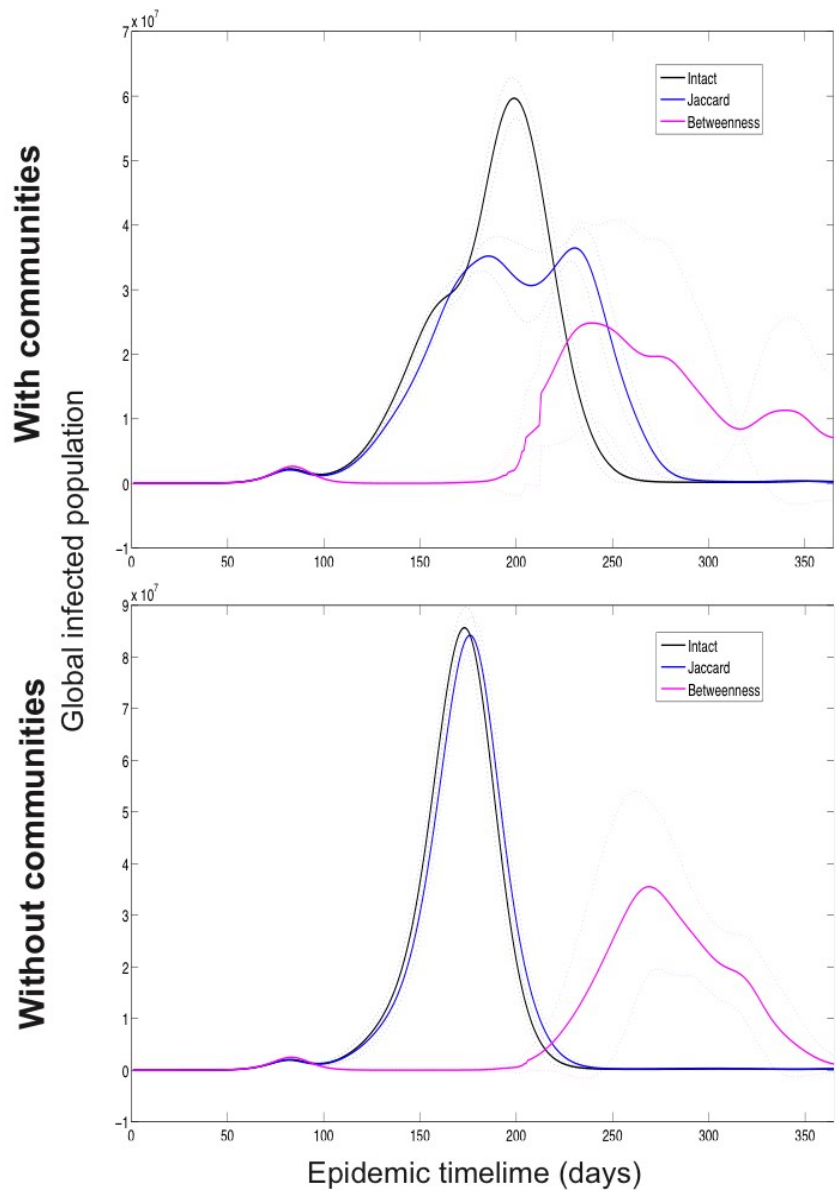

Worldwide infections over time following edge removal as selected by edge betweenness centrality and Jaccard coefficient. The two plots show results for two rewired versions of the network: one preserving only the original degree distribution and another preserving both the degree distribution plus original communities. Full lines show the averaged results of rewired versions of the airline network. Intact results are for the complete networks, while Jaccard and betweenness lines show the averaged results following the removal of 25% of edges by each respective measure. Dotted lines show each corresponding standard deviation.

This apparent advantage of betweenness even in networks without communities is due to its use of the capacity of each connection (edge weight), at

25% edge removal it will have removed most major high capacity connections from the network. Jaccard is a purely structural measure and without knowledge of capacity. The presence of communities is then critical for its performance. At lower levels of damage we see that Jaccard is better than edge betweenness centrality at reducing infected cases in networks with community structure.

# Discussion

Selecting specific edges for removal efficiently controls spreading in the airline network. Although this was not tested directly, cancelling fewer flights might also lead to fewer passengers that are affected by these policies compared to the approach of cancelling mostly flights from highly connected nodes (hubs). With the same number of removed connections, edge removal strategies resulted in both a larger slowdown of spreading and a resulting much smaller number of infected individuals compared to hub removal strategies.

Edge betweenness was best at predicting critical edges that carried the greater traffic weighted by number of passengers traveling resulting in a large reduction in infectious population; however we also observed that removing edges ranked using the purely structural Jaccard coefficient (see Fig. 2A) led to the greatest delay in reaching the peak of the epidemic. Among the best predictor edge measures, due to a computational complexity of $O(n^2)$, the Jaccard coefficient is the fastest measure to calculate, making it particularly suitable for large networks or networks where the topology frequently changes. Edge betweenness was the computationally most costly measure with $O(n*e)$, for a network with n nodes and e edges.

Whereas hub removal was the worst strategy in this study, node centrality might lead to better results. Indeed, previous findings [10] show that the most highly connected cities in the airline system do not necessarily have the highest node centrality. However, node centrality would be computationally as costly as edge betweenness.

Highly ranked connections predicted by edge measures were critical for the transmission of infections or activity and can be targeted individually with fewer disruptions for the overall network. In the transportation network studied, this means higher ranked individual connections could be cancelled instead of isolating whole cities from the rest of the world.

Results obtained from simulating the same spreading strategy over differently rewired versions of the airline network demonstrated the mechanism behind the performance of the Jaccard predictor in slowing down spreading in networks that display a community structure, as is the case for spatially distributed real-world networks [27][28][29]. This is a good measure for these types of networks, given its good computational efficiency and the little information it requires to compute the critical links - it needs nothing else than to know the connections between nodes.

The current study was testing different strategies and different percentages of removed edges leading to a large number of scenarios that had to be tested. Therefore, several simplifications had to be performed whose role could be investigated in future studies. First, only one starting point, Mexico City, for epidemics was tested. While this is in line with earlier studies using 1-3 starting points [5][23], it would be interesting to test whether there are exceptions to the outcomes presented here. Second, spreading was observed only in one season, summer. Previous work [23] has pointed out that the actual spreading pattern differs for different seasons. Third, only the 500 airports with the largest traffic volume rather than all 3,968 airports were included in the simulation. While this was done in order to be comparable with the earlier study of Hufnagel et al. [5], tests on the larger dataset would be interesting. Including airports with lower traffic volumes might preferable include national and regional airports within network modules. This could lead to a faster infection of regions; however, connections between communities would still remain crucial for the global spreading pattern.

Compared to our earlier study where the spreading of infection between airports rather than individuals was modelled [25], edge betweenness could reduce the maximally infected population number more than targeting network hubs. The Jaccard coefficient that showed very good performance in the earlier study [25], however, did not perform better than the hub strategy. The difference and product of node degrees were poor strategies for both spreading models. This indicates that metapopulation models can lead to a different evaluation of flight cancellation strategies for slowing down influenza spreading.

In conclusion, our results point to edge-based component removal for efficiently slowing spreading in airline and potentially other real-world networks.

# Materials and Methods

The network of connections between the top 500 airports is available under the resources link on our website http://www.biological-networks.org. Note that distribution of the complete data set, including all airports and traffic volumes, is not allowed due to copyright restrictions. However, the complete dataset can be purchased directly from OAG Worldwide Limited.

### Airline connections network

As in other work [5][10], we obtained scheduled flight data for one year provided by OAG Aviation Solutions (Luton, UK). This listed 1,341,615 records of worldwide flights operating from July 1, 2007 to July 30, 2008, which is estimated by OAG to cover 99% of commercial flights. The records include the cities of origin and destination, days of operation, and the type of aircraft in service for that route. Airports were uniquely identified by their IATA code together with their corresponding cities. These cities became the nodes in the network. Short-distance links corresponding to rail, boat, bus or limousine connections were removed from our data set. An edge connecting a pair of cities is present if at least one scheduled flight connected both airports. As in previous studies [5], we used a sub-graph containing the 500 top airports that was obtained by selecting the airports with greater seat traffic combining incoming and outgoing routes. This subset of airports still represents at least 95% of the global traffic, and as demonstrated in [30] it includes sufficient information to describe the global spread of influenza. We are allowed to make the restricted data set of 500 airports available and you can download it under the resources link at http://www.biological-networks.org/

### Spreading model

Our analysis is based on the stochastic equation-based (SEB) epidemic spreading model as used in [31], simulating the spreading of influenza both within cities and at a global level through flights connecting the cities' local airports. Within cities, a stochastically variable portion of the susceptible population establishes contact with infected individuals. This type of meta-population model accounts for 5 different states of individuals within cities: non-susceptible, susceptible, exposed, infectious, and removed (deceased). As we have not considered vaccination in this model we did not use the non-susceptible class in our study.

Movement of individuals between cities is determined deterministically from the daily average passenger seats on flights between cities. Once infectious, individuals will not travel. We have assumed a moderate level of transmissibility between individuals, where $R_0$ = 1.7, as also used in other influenza studies [31][32]. Note, however, that future epidemics of H5N1 and other viruses might have different In [5] a similar model including stochastic local dynamics was used, however it was focused on a specific outbreak of SARS (Severe Acute Respiratory Syndrome) and Hong Kong was considered its starting point.

### Edge removal strategies

Five candidate measures for predicting critical edges in networks were tested. The measures are based on range of different parameters including node similarity, degree and all pairs shortest paths. Measures are taken only once from the intact network and are not recomputed after each removal step.

Edge betweenness centrality [33][34] represents how many times that particular edge is part of the all-pairs shortest paths in the network. Edge betweenness can show the impact of a particular edge on the overall characteristic path length of the network; a high value reveals an edge that will quite likely increase the average number of steps needed for spreading.

The Jaccard similarity coefficient (or matching index [35][36]) shows how similar the neighbourhood connectivity structure of two nodes is, for example two nodes who shared the exact same set of neighbours would have the maximum similarity coefficient of 1. A low coefficient reveals a connection between two different network structures that might represent a "shortcut" between remote regions, making such low Jaccard coefficient edges a good target for removal.

The absolute difference of degrees for the adjacent nodes is another measure of similarity of two nodes. A large value here indicates a connection between a network hub a more sparsely connected region of the network.

The product of the degrees of the nodes connected by the edge is high when both nodes are highly connected (hubs).

For testing the absolute difference and product of degrees we also considered the opposite removal strategy (starting with lowest values) but the results showed to be consistently under-performing when compared to all other measures (not shown).

Finally, highly connected nodes will be detected and the nodes, and therefore all the edges of that node, will be removed from the network. Note that this is referred to as 'hub removal strategy' whereas the impact is shown in relation to the number of edges which are removed after each node removal.

### Simulation algorithm

Original simulation code, as used in [23], was obtained from the MIDAS Project (Research Computing Division, RTI International). The simulator was developed in Java (Sun Microsystems, USA) programming language using the AnyLogic$^{TM}$ (version 5.5, XJ Technologies, USA) simulation framework to implement the dynamical model. Network measures were implemented in custom MATLAB (R2008b, MathWorks, Inc., Natick, USA) code. Results were further processed in MATLAB . Simulations were run in parallel on a 16-core HP ProLiant server, using the Sun Java 6 virtual machine.

Edge betweenness centrality was implemented using the algorithm by Brandes [34]. Links between cities in the network were considered to be directed, the network used included a total of 24,009 edges.

Mexico City was used as a starting node as observed in in the recent 2009 H1N1 pandemic. The starting date of the epidemic was assumed to be 1 July, and the pandemic evolution is simulated over the following 365 days, covering all the effects of seasonality as seen in both the Southern and Northern hemispheres.

Following the removal of each group of edges ranked by each control strategy, the spreading simulations were repeated.

### Testing rewired networks

To test whether the mechanism of control arose from the particular community structure or degree distribution, we observed two different rewired versions of the original network. In one version only each individual node degree was maintained and the whole network was randomly rewired, destroying the original community structure. For the second, the original community structure was preserved but the sub-network within each community was rewired, so connections within the community were rearranged but the original inter- community links were preserved. Both rewiring strategies preserved the original degree structure by the commonly used algorithm [37] in order to maintain the same number of passengers departing from each city and the number of passengers is only shuffled to different destinations. This way both strategies did not change in the number of passengers departing from each city, only the connectivity structure was modified.

The original community structure was identified using an heuristic modularity optimization algorithm [38] which identified four distinct clusters. These are predominantly geographic: one for North and Central America, including Canada and Hawaii, another for South America, a third including the greater part of China (except Hong Kong, Macau and Beijing) and finally a fourth including all other airports (Fig. 1B).

Twenty rewired networks were generated for each version of the rewiring algorithm and the daily average evolution of influenza, using the same spreading algorithm as above, was taken across these 20 networks. This was repeated after the removal of each group of edges. Therefore each measure on each of the rewired lots combines 182,500 individual results.

# Acknowledgements

We thank http://www.FlightStats.com for providing location information for all airports and OAG Worldwide Limited for providing the worldwide flight data for one year.

# Funding information

Supported by WCU program through the National Research Foundation of Korea funded by the Ministry of Education, Science and Technology (R32-10142). Marcus Kaiser was also supported by the Royal Society (RG/2006/R2), the CARMEN e-science project (http://www.carmen.org.uk) funded by EPSRC (EP/E002331/1), and (EP/G03950X/1). Jose Marcelino was supported by EPSRC PhD studentship (CASE/CNA/06/25) with a contribution from e-Therapeutics plc.

# Competing interests

The authors have declared that no competing interests exist.

This knol is part of the collection: PLoS Currents: Influenza
(Recrudescent wave of pandemic A/H1N1... Next »)

## Comments

Comments are moderated, and will not be visible until one of the authors of this knol approves.

**Write New Comment ▼**